\documentclass[aps,twocolumn,pra,superscriptaddress,showpacs,tightenlines]{revtex4}
\usepackage{amsmath}
\usepackage{graphicx}
\usepackage{epstopdf}
\usepackage{epsfig}
\usepackage{amsfonts}

\begin{document}
%\begin{CJK*}{GBK}{song}
\title{Correlated two-photon scattering in cavity optomechanics}
\author{Jie-Qiao Liao}
\affiliation{Department of Physics and Institute of Theoretical Physics, The Chinese
University of Hong Kong, Shatin, Hong Kong Special Administrative Region,
People's Republic of China}
\affiliation{Center for Emergent Matter Science, RIKEN, Wako-shi, Saitama 351-0198, Japan}
\author{C. K. Law}
\affiliation{Department of Physics and Institute of Theoretical
Physics, The Chinese University of Hong Kong, Shatin, Hong Kong
Special Administrative Region, People's Republic of China}
\date{\today}

\begin{abstract}
We present an exact analytical solution of the two-photon
scattering in a cavity optomechanical system. This is achieved by
solving the quantum dynamics of the total system, including the
optomechanical cavity and the cavity-field environment, with the Laplace
transform method. The long-time solution reveals detailed physical
processes involved as well as the corresponding resonant photon
frequencies. We characterize the photon correlation induced in the scattering process
by calculating the two-photon joint spectrum of the long-time state.
Clear evidence for photon frequency anti-correlation can be observed in the joint spectrum.
In addition, we calculate the equal-time second-order correlation
function of the cavity photons. The results show that the radiation pressure coupling can induce photon blockade effect, which
is strongly modulated by the phonon sideband resonance. In particular,
we obtain an explicit expression of optomechanical coupling strength determining these
sideband modulation peaks based on the two-photon resonance condition.
\end{abstract}

%42.50.Pq Cavity quantum electrodynamics; micromasers
%42.50.Wk Mechanical effects of light on material media, microstructures and particles
%42.50.Ar Photon statistics and coherence theory
%07.10.Cm 07.10.Cm Micromechanical devices and systems

\maketitle
%\end{CJK*}

\section{Introduction}

The realization of photon-photon interaction at few-photon level has been
a research subject of major interest in quantum optics~\cite{Imamoglu1997,Kimble2005}. The significance of few-photon
interaction exists not only for studying the foundations of
quantum theory, but also for applications in quantum information science. Specifically, an important goal in the current field of research is the control of
two-photon correlations. Such a problem has been discussed in various nonlinear systems~\cite{Kojima2003,Fan2007,Shi2009,Liao2010,Roy2010,Zhang2011,Zheng2012}.
For example, it has been shown that photon-photon interaction can be achieved in
nonlinear Kerr media, and interesting quantum correlation effects
such as photon blockade~\cite{Imamoglu1997,Kimble2005,Houck2011} appear in
the strong-coupling regime. In particular, a strong Kerr nonlinearity can be achieved through the
interaction between light and atoms~\cite{Imamoglu1997,Kimble2005}.

Interestingly, an optomechanical cavity~\cite{Kippenberg2008,Marquardt2009,Favero2009}
driven by radiation-pressure can be mapped to a problem
with a Kerr-type interaction~\cite{Mancini1997,Bose1997,Gong2009}, and hence provides a
different class of systems to explore phenomena of interacting
photons through the control of mechanical motion of the cavity
mirrors. Indeed, Rabl~\cite{Rabl2011} has examined the photon
blockade effect in a continuously driven cavity optomechanical
system operated in the single-photon strong-coupling regime. In
such a regime, a single cavity photon can significantly change the
resonant frequency of the cavity field and leads to a range of
interesting effects, such as quantum state preparation~\cite{Mancini1997,Bose1997,Bouwmeester2003,Chen2011},
multiple mechanical sidebands~\cite{Nunnenkamp2011}, scattering~\cite{Liao2012},
single-photon cooling~\cite{cooling}, and optomechanical instability~\cite{Qian}. We note that several
recent experimental systems in cavity
optomechanics~\cite{Gupta2007,Brennecke2008,Eichenfield2009}
are approaching the single-photon strong-coupling regime.

In this paper, we investigate the quantum dynamics of a photon
pair interacting with a moving mirror in a cavity. Different from
previous studies~\cite{Rabl2011,Nunnenkamp2011} in which a
continuously driven system is treated perturbatively, our system
confined in the two-photon subspace is exactly solvable and
therefore provides a fundamental configuration to study how two
photons can become correlated via optomechanical coupling.
We shall focus on the scattering problem in which a
two-photon wave packet is injected into the cavity. By treating the
cavity field, the moving mirror, and the field continuum outside
the cavity as a whole system, we can solve the quantum state evolution
via the Wigner-Weisskopf method. From the long-time solution, we
identify quantitatively the probability amplitudes associated with
various scattering processes inside the cavity.

We emphasize that our exact solution provides a complete description of the full quantum
state, including the system's environment, and this is not directly attainable in previous related
studies based on the master equation of the system's reduced density matrix~\cite{Kronwald2012} or
approximated operator solutions in Heisenberg's picture~\cite{Rabl2011}. The knowledge of the full quantum
state enables us to learn about the details of underlying physical processes as well as
the quantum structure of interacting photons in optomechanical cavities.
In this paper, we also calculate the joint spectrum of the two scattered photons.
From the joint spectrum, we can see clear evidence for the photon frequency anti-correlation, which is a qualitative
evaluation of the two-photon correlation. In addition, we
calculate the equal-time second-order correlation function of the cavity photons at transient
times as a quantitative measurement of the two-photon correlation. The results indicate that photon blockade is strongly
modulated by the optomechanical coupling strength as well as the Franck-Condon factors involved. More importantly,
we find out the explicit modulation rule of the optomechanical coupling strength
based on the two-photon resonance.

\section{The Model}

\begin{figure}[tbp]
\center
\includegraphics[bb=108 381 392 722, width=3.2 in]{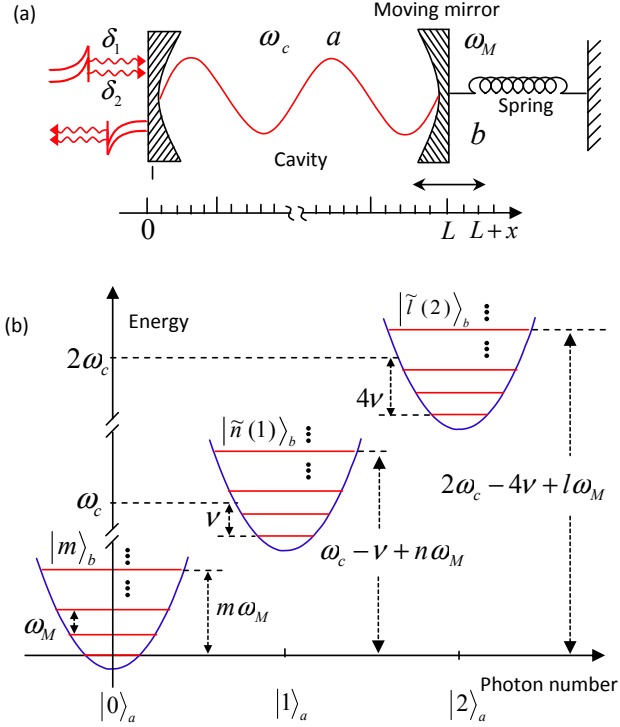}
\caption{(Color online) (a) Schematic diagram of a
Fabry-P\'{e}rot-type optomechanical cavity formed by a fixed end
mirror and a moving end mirror. (b) The energy-level structure
(unscaled) of the optomechanical cavity (limited in the zero-,
one-, and two-photon subspaces).} \label{setup}
\end{figure}
The system under consideration is a Fabry-P\'{e}rot-type
optomechanical cavity formed by a fixed end mirror and a moving
end mirror [see Fig.~\ref{setup}(a)]. We focus on a single-mode
cavity field, which is coupled to the mechanical oscillation of
the moving mirror via radiation pressure. The moving end mirror is
assumed to be perfect and the fixed one is partially transparent.
In a rotating frame defined by the unitary transformation $e^{-i
N \omega_c t}$ with $N=a^{\dagger}a+\int_{0}^{\infty}c_{k}^{\dagger}c_{k}dk$ being the
total photon number operator, the Hamiltonian of the whole system
including the optomechanical cavity and the environment reads as
\begin{eqnarray}
H_{I}&=&\hbar\omega_{M}b^{\dagger}b-\hbar g_{0}a^{\dagger}a(b^{\dagger}+b)
+\int_{0}^{\infty}\hbar\Delta_{k}c_{k}^{\dagger}c_{k}dk\nonumber\\
&&+\hbar\xi\int_{0}^{\infty}(c_{k}^{\dagger}a+a^{\dagger}c_{k})dk. \label{hamiltonian}
\end{eqnarray}
Here $a$ $(a^{\dagger})$ and $b$ $(b^{\dagger })$ are,
respectively, the annihilation (creation) operators of the cavity
field and the moving mirror, with the respective frequencies
$\omega_{c}$ and $\omega_{M}$. $c_{k}$ $(c_{k}^{\dagger })$ is
the annihilation (creation) operator of the continuous field
mode $k$ outside the cavity with the resonant frequency $\omega
_{k}$ ($\Delta _{k}=\omega _{k}-\omega _{c}$ is the detuning). The
radiation-pressure coupling appearing in the second term of
$H_{I}$ is characterized by the coupling strength $g_{0}$, and the
coupling between the cavity field and the outside fields is modeled
by the hopping interaction with the coupling strength
$\xi=\sqrt{\gamma_{c}/2\pi}$ ($\gamma_{c}$ is the cavity-field
decay rate). Since the decay rate of the mechanical resonator
$\gamma_M$ can be much smaller than $\gamma_{c}$, we will neglect
the dissipation of the mechanical resonator in our discussions. This is justified as
long as the scattering processes are completed in a time much
shorter than $\gamma_M^{-1}$.

In this paper, we will focus on the two-photon scattering problem.
Because the total photon number is a conserved quantity, the
Hilbert space for the photon part in this problem is spanned by
three types of basis vectors:
$\vert2\rangle_{a}\vert\emptyset\rangle$,
$\vert1\rangle_{a}\vert1_{k}\rangle$, and
$\vert0\rangle_{a}\vert1_{p},1_{q}\rangle$, where
$\vert2\rangle_{a}\vert\emptyset\rangle$ stands for the
state with two photons in the cavity and the outside fields are in
a vacuum, $\vert1\rangle_{a}\vert1_{k}\rangle$ represents the
state with one photon in the cavity, one photon in the $k$ mode
of the outside fields,  and the state
$\vert0\rangle_{a}\vert1_{p},1_{q}\rangle$ means that there is no
photon in the cavity, two photons are in the $p$ and $q$ modes of the
outside fields. By using these basis vectors for the fields, the
state vector of the total system at time $t$ is denoted by
\begin{eqnarray}
\vert\Phi(t)\rangle&=&\sum_{m=0}^{\infty}A_{m}(t)\vert
2\rangle_{a}\vert\emptyset\rangle \vert\tilde{m}(2)\rangle_{b}\nonumber\\
&&+\sum_{m=0}^{\infty}\int_{0}^{\infty}dkB_{m,k}(t)\vert
1\rangle_{a} \vert1_{k}\rangle \vert\tilde{m}(1)\rangle_{b} \nonumber\\
&&+\sum_{m=0}^{\infty}\int_{0}^{\infty}dp\int_{0}^{p}dq
C_{m,p,q}(t)\vert0\rangle_{a}\vert
1_{p},1_{q}\rangle\vert m\rangle_{b},\label{generalstate}
\end{eqnarray}
where we have introduced the single- and two-photon displaced number
states for the moving mirror
\begin{eqnarray}
|\tilde{m}(1)\rangle_{b}&\equiv&
e^{\beta_{0}(b^{\dagger}-b)}|m\rangle_{b},\nonumber\\
|\tilde{m}(2)\rangle_{b}&\equiv&
e^{2\beta_{0}(b^{\dagger}-b)}|m\rangle_{b},\label{dispnumsta}
\end{eqnarray}
with $\beta_{0}\equiv g_{0}/\omega_{M}$ being the single-photon
displacement quantity. We note that these displaced number states
are eigenstates of the Hamiltonian
\begin{equation}
H_{\textrm{opc}}=\hbar\omega_{c}a^{\dagger}a+\hbar\omega_{M}b^{\dagger}b-\hbar
g_{0}a^{\dagger}a(b^{\dagger}+b)
\end{equation}
of the optomechanical cavity, as defined by the eigen-equation
\begin{equation}
H_{\textrm{opc}}|l\rangle_{a}|\tilde{m}(l)\rangle_{b}
=\hbar(l\omega_{c}+m\omega_{M}-l^{2}\nu)|l\rangle_{a}|\tilde{m}(l)\rangle_{b},
\end{equation}
where $\nu\equiv g_{0}^{2}/\omega_{M}$ is the single-photon-state frequency shift.
The energy-level structure limited in the zero-, one-,
and two-photon subspaces of the optomechanical cavity is shown in Fig.~\ref{setup}(b).

Based on Eqs.~(\ref{hamiltonian}), (\ref{generalstate}), and the
Schr\"{o}dinger equation, we obtain the following equations of
motion for probability amplitudes
\begin{widetext}
\begin{subequations}
\label{equationofmotion}
\begin{gather}
\dot{A}_{m}(t)=-i(m\omega_{M}-4\nu)A_{m}(t)-i\sqrt{2}\xi
\sum_{n=0}^{\infty}\int_{0}^{\infty}\langle\tilde{m}(2)\vert_{b}\vert\tilde{n}(1)
\rangle_{b}B_{n,k}(t)dk,\\
\dot{B}_{m,k}(t)=-i(\Delta_{k}-\nu+m\omega_{M})B_{m,k}(t)-i\sqrt{2}\xi\sum_{n=0}^{\infty}\langle\tilde{m}(1)\vert_{b}
\vert\tilde{n}(2)\rangle_{b}A_{n}(t)-i\xi\sum_{n=0}^{\infty}\int_{0}^{\infty}\langle\tilde{m}(1)\vert_{b}\vert
n\rangle_{b}C_{n,p,k}(t)dp,\\
\dot{C}_{m,p,q}(t)=-i(\Delta_{p}+\Delta_{q}+m\omega_{M})
C_{m,p,q}(t)-i\xi\sum_{n=0}^{\infty}\langle
m\vert_{b}\vert\tilde{n}(1)\rangle _{b}[B_{n,p}(t)+B_{n,q}(t)].
\end{gather}
\end{subequations}
\end{widetext}
Here we point out that the transition rates associated with photon scattering processes
are determined by the Franck-Condon factors $\langle\tilde{m}(2)\vert_{b}\vert\tilde{n}(1)
\rangle_{b}$, $\langle\tilde{m}(1)\vert_{b}
\vert\tilde{n}(2)\rangle_{b}$, $\langle\tilde{m}(1)\vert_{b}\vert
n\rangle_{b}$, and $\langle
m\vert_{b}\vert\tilde{n}(1)\rangle _{b}$, which can be calculated based on the
relation~\cite{Oliveira1990}
\begin{eqnarray}
&&\langle m\vert_{b}e^{\beta_{0}(b^{\dagger}-b)}\vert n\rangle_{b}\nonumber\\
&&=\left\{
\begin{array}{c}
\sqrt{\frac{m!}{n!}}e^{-\frac{\beta_{0}^{2}}{2}}
(-\beta_{0})^{n-m}L_{m}^{n-m}(\beta_{0}^{2}),\hspace{0.1 cm}n\geq m,\\
\sqrt{\frac{n!}{m!}}e^{-\frac{\beta_{0}^{2}}{2}}
\beta_{0}^{m-n}L_{n}^{m-n}(\beta_{0}^{2}),\hspace{0.1 cm}m>n,
\end{array}
\right.
\end{eqnarray}
where $L_{r}^{s}(x)$ is the associated Laguerre polynomial.

\section{Two-photon scattering solution}

Initially the cavity field is in the vacuum state and two photons
are injected into the cavity. These two incident photons are
prepared in a wave packet form  outside the cavity. To facilitate
analytic treatment by the Laplace transform, we examine the two-photon
wave packet with Lorentzian spectrum. Without loss of generality,
the initial state of the mirror is assumed to be number state $\vert
n_{0}\rangle_{b}$. Once the solution in this case is found, the
solution for general initial mirror states can be obtained
accordingly by superposition. Explicitly, the initial condition is
specified by $A_{m}(0)=0$, $B_{m,k}(0)=0$, and
\begin{eqnarray}
C_{m,p,q}(0)&=&\left[\frac{\mathcal{N}\delta_{m,n_{0}}}{(\Delta
_{p}-\delta_{1}+i\epsilon)(\Delta_{q}-\delta_{2}+i\epsilon)}+\delta_{1}\leftrightarrow\delta_{2}\right],
\label{scainicondition}\nonumber\\
\end{eqnarray}
where $\delta_{j=1,2}=\omega_{j}-\omega_{c}$ ($\omega_{j}$ is the resonant frequency) and $\epsilon$ define the center
detuning and spectral width of the two photons.
The normalization constant $\mathcal{N}$ is
\begin{eqnarray}
\mathcal{N}=\frac{\epsilon
}{\pi}\left(1+\frac{4\epsilon^{2}}{(\delta_{1}-\delta_{2})
^{2}+(2\epsilon)^{2}}\right)^{-1/2}.
\end{eqnarray}

By using the Laplace transform, analytical solution of Eq.~(\ref{equationofmotion}) subjected to the initial condition
can be found (see Appendix~\ref{apptwophoscat}). In the long-time
limit when the scattering is completed and the two photons exit
the cavity, the solution is given by: $A_{n_{0},m}(\infty)=0$,
$B_{n_{0},m,k}(\infty)=0$, and
\begin{eqnarray}
C_{n_{0},m,p,q}(\infty)&=&\mathcal{N}[(C_{\textrm{I}}+C_{\textrm{II}}+C_{\textrm{III}}+C_{\textrm{IV}})\nonumber\\
&&+(\Delta _{p}\leftrightarrow\Delta_{q})]e^{-i(\Delta
_{p}+\Delta_{q}+m\omega_{M})t}.\label{sctlontsolu}
\end{eqnarray}
Here, we add the subscript $n_{0}$ in $A_{n_{0},m}(\infty)$,
$B_{n_{0},m,k}(\infty)$, and $C_{n_{0},m,p,q}(\infty)$ to mark the
mirror's initial state $|n_{0}\rangle_{b}$. The four transition amplitude components $C_{\textrm{I}}$,
$C_{\textrm{II}}$, $C_{\textrm{III}}$, and $C_{\textrm{IV}}$ are
given by
\begin{eqnarray}
C_{\textrm{I}}&=&\frac{1}{(\Delta_{p}-\delta_{1}+i\epsilon)}\frac{1}{(\Delta_{q}-\delta_{2}+i\epsilon)}\delta_{m,n_{0}},\nonumber\\
C_{\textrm{II}}&=&\sum_{n=0}^{\infty}\frac{-i\gamma_{c}F_{\textrm{II}}}{M_{1}M_{2}(\Delta_{q}-\delta_{2}+i\epsilon)}
+\delta_{1}\leftrightarrow\delta_{2},\nonumber\\
C_{\textrm{III}}&=&\sum_{n,n^{\prime},l=0}^{\infty}\frac{-\gamma_{c}^{2}F_{\textrm{III}}}{M_{1}M_{3}M_{4}M_{5}}
+\delta_{1}\leftrightarrow\delta_{2},\nonumber\\
C_{\textrm{IV}}&=&\sum_{n,n^{\prime},l=0}^{\infty}\frac{-2\gamma_{c}^{2}F_{\textrm{IV}}}{M_{1}M_{3}M_{4}M_{6}}+\delta_{1}\leftrightarrow\delta_{2}.
\label{ci-civ}
\end{eqnarray}
where we have introduced $F_{\textrm{II}-\textrm{IV}}$ to denote products of Franck-Condon factors
\begin{eqnarray}
F_{\textrm{II}}&=&\langle m\vert_{b}\vert\tilde{n}(1)\rangle_{b}\langle\tilde{n}(1)\vert_{b}\vert n_{0}\rangle_{b},\nonumber\\
F_{\textrm{III}}&=&\langle
m\vert_{b}\vert\tilde{n}(1)\rangle_{b}\langle\tilde{n}(1)\vert
_{b}\vert n^{\prime}\rangle_{b}\langle n^{\prime}\vert_{b}\vert
\tilde{l}(1)\rangle_{b}\langle\tilde{l}(1)\vert_{b}\vert
n_{0}\rangle_{b},\nonumber\\
F_{\textrm{IV}}&=&\langle m\vert _{b}\vert \tilde{n}(1)\rangle_{b}\langle
\tilde{n}(1)\vert_{b}\vert\tilde{n}^{\prime}(2)\rangle_{b}\langle
\tilde{n}^{\prime}(2)\vert_{b}\vert\tilde{l}(1)\rangle_{b}\langle
\tilde{l}(1)\vert_{b}\vert n_{0}\rangle_{b}.\label{FCfactors}\nonumber\\
\end{eqnarray}
In addition, the denominators in Eq.~(\ref{ci-civ}) are defined by
\begin{eqnarray}
M_{1}&=&\Delta_{p}+\nu+(m-n)\omega_{M}+i\frac{\gamma_{c}}{2},\nonumber\\
M_{2}&=&\Delta_{p}-\delta_{1}+(m-n_{0})\omega_{M}+i\epsilon,\nonumber\\
M_{3}&=&\Delta_{p}+\Delta_{q}-\delta_{1}-\delta_{2}+(m-n_{0})\omega_{M}+2i\epsilon,\nonumber\\
M_{4}&=&\Delta_{p}+\Delta_{q}-\delta_{1}+\nu+(m-l)\omega_{M}+i\frac{\gamma_{c}}{2}+i\epsilon,\nonumber\\
M_{5}&=&\Delta_{p}-\delta_{1}+(m-n^{\prime})\omega_{M}+i\epsilon,\nonumber\\
M_{6}&=&\Delta_{p}+\Delta_{q}+4\nu+(m-n^{\prime})\omega_{M}+i\gamma_{c}.\label{m1-m6}
\end{eqnarray}
From Eq.~(\ref{m1-m6}) we are able to determine the resonance conditions involved in the photon scattering process.

%%%%%%%%%%%%%%%%%%%%%%%%%%%%%%%%%%%%%%%%%%%%%%%%%%%%%%%%%%%%%%%%%%%%%%%%
\begin{figure*}[tbp]
\center
\includegraphics[bb=71 466 554 733, width=7 in]{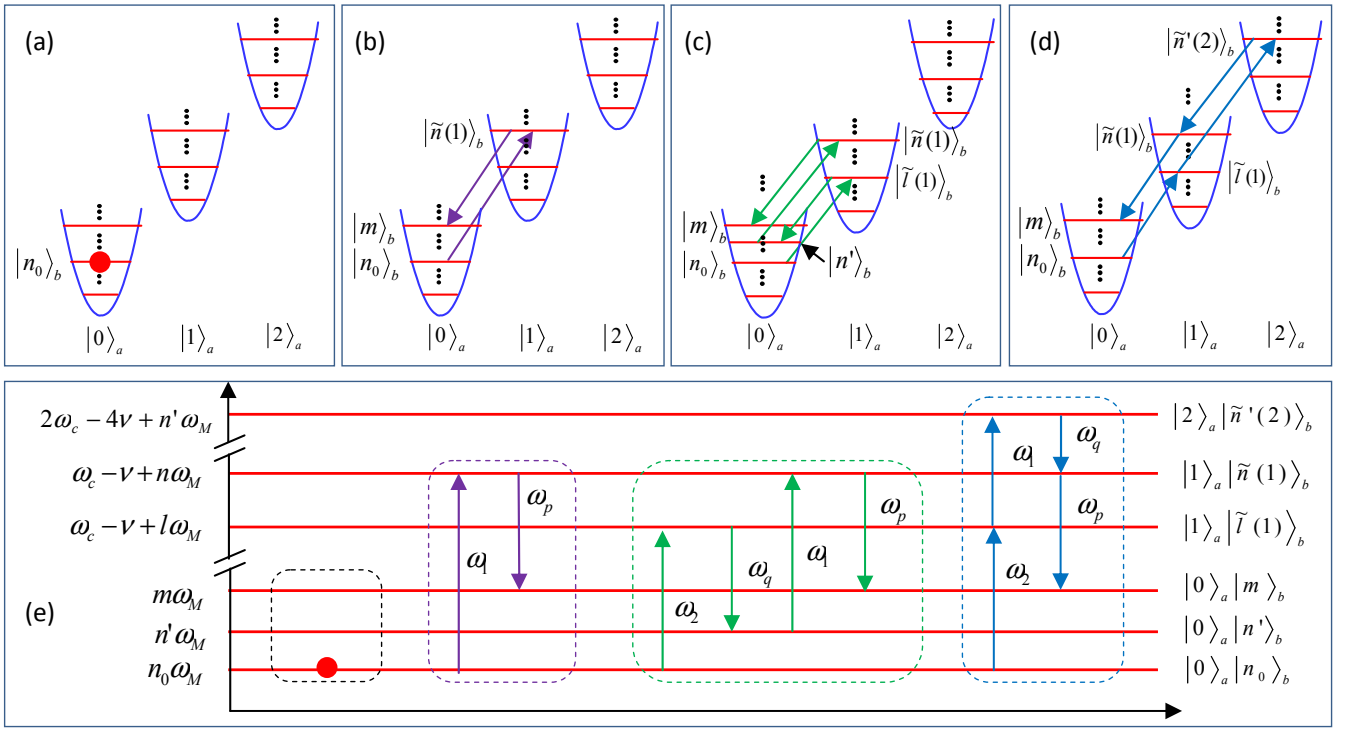}
\caption{(Color online) (a-d) Four types of transitions from
states $|0\rangle_{a}|n_{0}\rangle_{b}$ to
$|0\rangle_{a}|m\rangle_{b}$ of the optomechanical cavity in the
two-photon scattering process. (e) The eigen-energies of these
states involved in the four transition processes.}
\label{transitions}
\end{figure*}
%%%%%%%%%%%%%%%%%%%%%%%%%%%%%%%%%%%%%%%%%%%%%%%%%%%%%%%%%%%%%%%%%%%%%%%%
We point out that the four amplitudes
$C_{\textrm{I}-\textrm{IV}}$ correspond to four different
physical processes in the two-photon scattering (see
Fig.~\ref{transitions}). The state transitions associated with
these processes can be identified by $F_{\textrm{II}-\textrm{IV}}$.
For the term $C_{\textrm{I}}$, it describes a direct two-photon
reflection by the fixed end mirror without entering the cavity.
Therefore the mirror state does not change, and the two reflected
photons remain in the Lorentzian wave packet state [Fig.~\ref{transitions}(a)].

The term $C_{\textrm{II}}$ corresponds to the one-photon
scattering and one-photon reflection process,  i.e., just one
photon enters the cavity and the other photon is reflected by the
fixed end mirror. The entered photon induces the transition
process $|0\rangle_{a}\vert
n_{0}\rangle_{b}\rightarrow|1\rangle_{a}\vert\tilde{n}(1)\rangle_{b}\rightarrow
|0\rangle_{a}|m\rangle_{b}$ [Fig.~\ref{transitions}(b)], and such
a process has the resonance conditions:
$n_{0}\omega_{M}+\omega_{1}=\omega_{c}-\nu+n\omega_{M}$ and
$\omega_{p}=(\omega_{c}-\nu+n\omega_{M})-m\omega_{M}$ [Fig.~\ref{transitions}(e)], according to
the poles of $C_{\textrm{II}}$, i.e., $\textrm{Re}(M_{1})=0$ and $\textrm{Re}(M_{1}-M_{2})=0$.

The third term $C_{\textrm{III}}$ can be interpreted as a
sequential two-photon scattering process, in which the second
photon enters the cavity after the first photon emitted out of the
cavity. Such an interpretation is obtained from $F_{\textrm{III}}$ that the maximum cavity photon number is one.
In this case, the system experiences the following transitions
$|0\rangle_{a}\vert n_{0}\rangle_{b}\rightarrow|1\rangle_{a}\vert
\tilde{l}(1)\rangle_{b}\rightarrow|0\rangle_{a}\vert
n^{\prime}\rangle_{b}\rightarrow|1\rangle_{a}|\tilde{n}(1)\rangle_{b}\rightarrow|0\rangle_{a}\vert
m\rangle_{b}$ [Fig.~\ref{transitions}(c)]. As depicted in
Fig.~\ref{transitions}(e), the photon excitation processes are
governed by
$n_{0}\omega_{M}+\omega_{2}=\omega_{c}-\nu+l\omega_{M}$ [i.e., $\textrm{Re}(M_{4}-M_{3})=0$] and
$n'\omega_{M}+\omega_{1}=\omega_{c}-\nu+n\omega_{M}$ [$\textrm{Re}(M_{1}-M_{5})=0$], and the
frequencies of the two resonant emitted photons are
$\omega_{q}=(\omega_{c}-\nu+l\omega_{M})-n'\omega_{M}$ [$\textrm{Re}(M_{4}-M_{5})=0$] and
$\omega_{p}=(\omega_{c}-\nu+n\omega_{M})-m\omega_{M}$ [$\textrm{Re}(M_{1})=0$].

The fourth term $C_{\textrm{IV}}$ corresponds to a `genuine' two-photon process involving the interaction with two photons inside
the cavity. This is revealed in $F_{\textrm{IV}}$
having the states with two cavity photons. The is a kind of
two-photon cascade scattering in which the system experiences the
transitions $|0\rangle_{a}\vert
n_{0}\rangle_{b}\rightarrow|1\rangle_{a}\vert
\tilde{l}(1)\rangle_{b}\rightarrow|2\rangle_{a}\vert
\tilde{n}^{\prime}(2)\rangle_{b}\rightarrow|1\rangle_{a}|\tilde{n}(1)\rangle_{b}\rightarrow|0\rangle_{a}\vert
m\rangle_{b}$ [Fig.~\ref{transitions}(d)]. When the two photons are in the cavity,
the mirror will experience an energy shift $-4\nu$. This extra energy shift (and the phonon sidebands, as we will show below) provides the
physical mechanism to create the photon blockade in the cavity. Due to the two-cavity-photon state is involved, the resonance conditions for
the photon excitation processes are $n_{0}\omega_{M}+\omega_{2}=\omega_{c}-\nu+l\omega_{M}$ [$\textrm{Re}(M_{4}-M_{3})=0$]
and $(\omega_{c}-\nu+l\omega_{M})+\omega_{1}=2\omega_{c}-4\nu+n'\omega_{M}$ [$\textrm{Re}(M_{6}-M_{4})=0$].
In addition, the frequencies of the two emitted photons
are $\omega_{q}=(2\omega_{c}-4\nu+n'\omega_{M})-(\omega_{c}-\nu+n\omega_{M})$ [$\textrm{Re}(M_{6}-M_{1})=0$] and
$\omega_{p}=(\omega_{c}-\nu+n\omega_{M})-m\omega_{M}$ [$\textrm{Re}(M_{1})=0$]. These resonance conditions can also be seen from Fig.~\ref{transitions}(e).

The above analysis expatiated the physical picture for creation of two-photon correlation.
In fact, after the scattering, the two photons will also entangle with the mirror. We can see this point by examining the
scattering state of the system. We know from Eqs.~(\ref{generalstate}) and (\ref{sctlontsolu}) that,
corresponding to the mirror's initial state $\vert
n_{0}\rangle_{b}$, the long-time state of the system is $\vert
0\rangle_{a}\otimes\vert\Phi_{n_{0}}(\infty)\rangle$ with
\begin{eqnarray}
\vert\Phi_{n_{0}}(\infty)\rangle=\sum_{m=0}^{\infty
}\int_{0}^{\infty }dp\int_{0}^{p}dqC_{n_{0},m,p,q}(\infty)\vert
m\rangle_{b}\vert1_{p},1_{q}\rangle.\label{emipurlonsta}\nonumber\\
\end{eqnarray}
Physically, the cavity is in a vacuum in the long-time limit,
and hence it decouples with the environment and the mirror. The state $\vert\Phi_{n_{0}}(\infty)\rangle$
is an entangled state involving these outside fields and
the mirror. We can understand this entanglement from the dependence
of the emitted photon modes on the mirror's final state
$|m\rangle_{b}$. In the scattering process, the modes of the two
emitted photons depend on the final state of the mirror [see Fig.~\ref{transitions}(e)].

\section{Two-photon joint spectrum}

We know from the above section that the radiation-pressure coupling can induce photon correlation.
We now show how to indicate qualitatively this correlation from the joint spectrum of the two scattered photons.
Corresponding to the cases where the mirror is initially in a pure state
$|\varphi\rangle_{b}=\sum_{n_{0}=0}^{\infty}c_{n_{0}}|n_{0}\rangle_{b}$ and
a mixed state $\rho_{b}=\sum_{n_{0}=0}^{\infty}p_{n_{0}}|n_{0}\rangle_{b}\langle n_{0}|_{b}$,
the joint spectrum functions~\cite{Liao2010} are defined by
\begin{subequations}
\begin{gather}
S(\Delta_{p},\Delta_{q})=\sum_{m=0}^{\infty}\left\vert
\sum_{n_{0}=0}^{\infty}c_{n_{0}}C_{n_{0},m,p,q}(\infty)\right\vert^{2},\label{spectpure}\\
S(\Delta_{p},\Delta_{q})=\sum_{m=0}^{\infty}\sum_{n_{0}=0}^{\infty}
p_{n_{0}}\vert C_{n_{0},m,p,q}(\infty)\vert^{2}.\label{spectmixed}
\end{gather}
\end{subequations}
%%%%%%%%%%%%%%%%%%%%%%%%%%%%%%%%%%%%%%%%%%%%%%%%%%%%
\begin{figure}[tbp]
\center
\includegraphics[bb=51 440 412 759, width=\columnwidth]{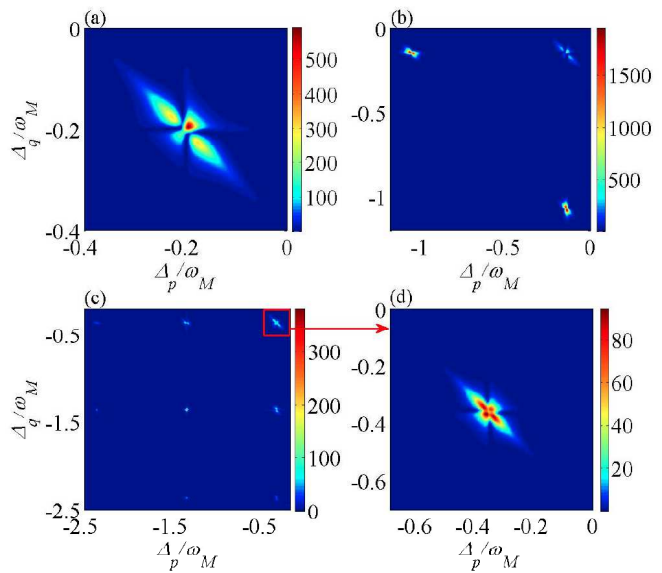}
\caption{(Color online) Plot of the two-photon joint spectrum $S(\Delta_{p},\Delta_{q})$ for the ground state $|0\rangle_{b}$ case.
For a small $\beta_{0}\ll1$, we plot the spectrum by expanding the $C_{n_{0},m,p,q}(\infty)$ up to (a) the zero-order ($\nu=0.2$) and (b) the first-order ($\beta_{0}=0.4$) of $\beta_{0}$.
(c) For a large $\beta_{0}$, the joint spectrum includes many sidebands ($\beta_{0}=0.6$). (d) The zoomed view of the peak
with the center position $\Delta_{p}/\omega_{M}=\Delta_{q}/\omega_{M}=-0.36$ from the subfigure (c). Other parameters are:
$\gamma_{c}/\omega_{M}=0.1$, $\epsilon/\omega_{M}=0.01$, and $\delta_{1}=\delta_{2}=-\nu$.}\label{jointspect}
\end{figure}
%%%%%%%%%%%%%%%%%%%%%%%%%%%%%%%%%%%%%%%%%%%%%%%%%%%%

In Fig.~\ref{jointspect}, we plot the two-photon joint spectrum $S(\Delta_{p},\Delta_{q})$ as a
function of the frequencies $\Delta_{p}$ and $\Delta_{q}$ when the initial state of the mirror is
$|0\rangle_{b}$. We firstly consider the case of
$\beta_{0}\ll1$ so that it is reasonable to approximately expand the probability amplitude $C_{n_{0},m,p,q}(\infty)$
up to the zero- and first-orders of $\beta_{0}$, i.e., using the approximations
\begin{eqnarray}
\langle m\vert_{b}e^{\beta_{0}(b^{\dagger}-b)}\vert n\rangle_{b}&\approx&\delta_{m,n},\nonumber\\
\langle m\vert_{b}e^{\beta_{0}(b^{\dagger}-b)}\vert n\rangle_{b}&\approx&\delta_{m,n}+\beta_{0}(\sqrt{n+1}\delta_{m,n+1}-\sqrt{n}\delta_{m,n-1})\nonumber\\
\end{eqnarray}
in the zero- and first-order expansions, respectively. We note that the energy shift terms
$\nu$ and $4\nu$ in Eq.~(\ref{m1-m6}) are kept because $\nu=g_{0}\beta_{0}$ could be larger than $\gamma_c$ even for small $\beta_{0}$.
Actually, the Kerr nonlinear energy shift $\nu(a^{\dagger}a)^{2}$ is responsible for the generation of photon correlation.
For the zero-order expansion, the effect of the sidebands is completely canceled, and the present optomechanical
system reduces to a Kerr nonlinear cavity with the Kerr parameter $\nu$. When $\nu>\gamma_c$, two free photons scattered by
the Kerr nonlinear cavity will be correlated~\cite{Liao2010}. This can be seen from the joint spectrum of the two photons [Fig.~\ref{jointspect}(a)].
The two scattered photons are frequency anticorrelated with a probability concentrated along the line parallel to $\Delta_{p}+\Delta_{q}=0$.
For larger $\beta_{0}$, the phonon sidebands will be involved in the spectrum.
In the first-order expansion case, we can see two sideband peaks in the two directions of the joint spectrum.
With the increasing of $\beta_{0}$, more and more sidebands can be observed in the spectrum. In Fig.~\ref{jointspect}(c),
we plot the joint spectrum in the single-photon strong coupling regime $\beta_{0}=0.6$. From the spectrum,
we can see that the pattern of these sideband peaks is concentrated along the line parallel to $\Delta_{p}+\Delta_{q}=0$.
This point can be seen clearly from Fig.~\ref{jointspect}(d), which is a zoomed view of the peak with the center located at $\Delta_{p}=\Delta_{q}=-\nu$ in Fig.~\ref{jointspect}(c).
We should emphasize that the joint spectrum is experimentally measurable by detecting the probability distribution
of the two scattered photons.

\section{Photon blockade in cavity}

In the previous section, we have studied the two-photon joint spectrum based on the long-time scattering solution.
Actually, we can quantitatively study the photon correlation by calculating the second-order correlation function for cavity photons.
However, we now need to consider the transient dynamics rather than the long-time dynamics of the system,
because there are no photons in the cavity in the long-time limit. In the following, we calculate the transient
dynamics of the system based on the solution given in the Appendix. Corresponding to these two cases of initial states $|\varphi\rangle_{b}$
and $\rho_{b}$, the probabilities for finding one and
two photons in the cavity can be obtained as
\begin{eqnarray}
P_{1}(t)&=&\sum_{m=0}^{\infty}\int_{0}^{\infty}dk\left\vert\sum_{n_{0}=0}^{\infty}c_{n_{0}}
B_{n_{0},m,k}(t)\right\vert^{2},\nonumber\\
P_{2}(t)&=&\sum_{m=0}^{\infty}\left\vert\sum_{n_{0}=0}^{\infty}c_{n_{0}}
A_{n_{0},m}(t)\right\vert^{2},\label{pureprobabilits}
\end{eqnarray}
and
\begin{eqnarray}
P_{1}(t)&=&\sum_{n_{0}=0}^{\infty}p_{n_{0}}\sum_{m=0}^{\infty}\int_{0}^{\infty}dk\left\vert
B_{n_{0},m,k}(t)\right\vert^{2},\nonumber\\
P_{2}(t)&=&\sum_{n_{0}=0}^{\infty}p_{n_{0}}\sum_{m=0}^{\infty}\left\vert
A_{n_{0},m}(t)\right\vert^{2},\label{mixedprobabilits}
\end{eqnarray}
where $A_{n_{0},m}(t)$ and $B_{n_{0},m,k}(t)$ are obtained by
making the inverse Laplace transform on Eqs.~(\ref{srepofakscat})
and~(\ref{srepofbmkscat}).
%%%%%%%%%%%%%%%%%%%%%%%%%%%%%%%%%%%%%%%%%%%%%%%%%%%%
\begin{figure}[tbp]
\center
\includegraphics[bb=3 46 397 522, width=\columnwidth]{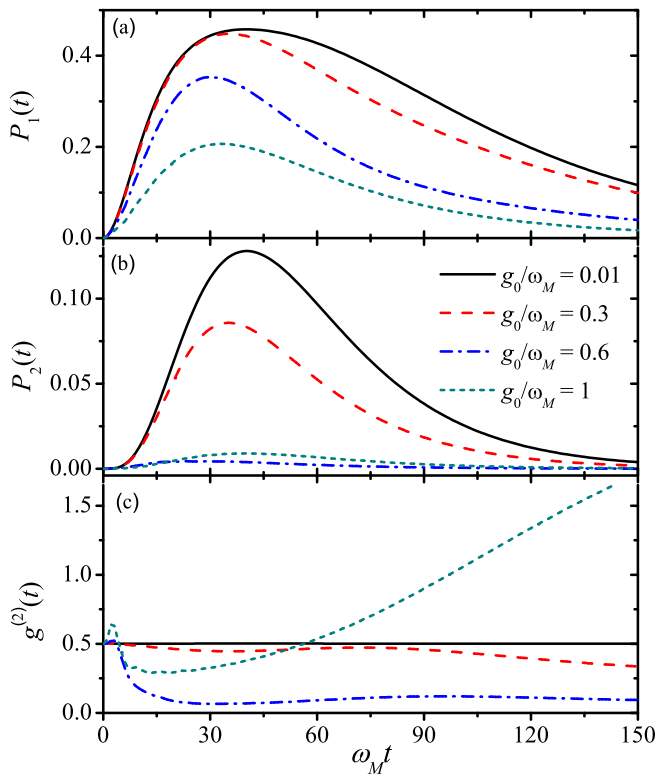}
\caption{(Color online) Evolution of (a) the single-photon
probability $P_{1}(t)$ and (b) two-photon probability $P_{2}(t)$ in the cavity for various
$g_{0}$. (c) Equal-time second-order correlation function $g^{(2)}(t)$ vs the scaled time $\omega_{M}t$. Other parameters are:
$\gamma_{c}/\omega_{M}=0.1$, $\epsilon/\omega_{M}=0.01$,
$c_{n_{0}}=\delta_{n_{0},0}$, and $\delta_{1}=\delta_{2}=-\nu$.}\label{sinphoresonce}
\end{figure}
%%%%%%%%%%%%%%%%%%%%%%%%%%%%%%%%%%%%%%%%%%%%%%%%%%%%

For the cavity field, the equal-time second-order correlation
function is defined by~\cite{London2000}
\begin{equation}
g^{(2)}(t)\equiv\frac{\textrm{Tr}[\rho(t) a^{\dagger}a^{\dagger}aa]}{(\textrm{Tr}[\rho(t) a^{\dagger}a])^{2}},\label{equtimcorrefunt}
\end{equation}
where $\rho(t)$ is the density matrix of the total system at time
$t$. In terms of Eqs.~(\ref{generalstate}), (\ref{pureprobabilits}), and (\ref{mixedprobabilits}),
Eq.~(\ref{equtimcorrefunt}) can be expressed as
\begin{equation}
g^{(2)}(t)=\frac{2P_{2}(t)}{[2P_{2}(t)+P_{1}(t)]^{2}}.\label{g2ofprobab}
\end{equation}
Photon blockade effect corresponds to $g^{(2)}(t)\ll1$~\cite{Werner1999}, and the
effect is measured by how small $g^{(2)}(t)$ is.

In what follows, we will study how the cavity photon statistics
depends on the radiation-pressure coupling strength $g_{0}$. In
Fig.~\ref{sinphoresonce}, we plot the time evolution of
$P_{1}(t)$, $P_{2}(t)$, and $g^{(2)}(t)$ for various $g_{0}$ at
$\delta_{1}=\delta_{2}=-\nu$, which corresponds to a single photon
resonance [see Fig.~\ref{setup}(b)]. Here we assume that the initial state of the mirror
is $|0\rangle_{b}$, i.e., $c_{n_{0}}=\delta_{n_{0},0}$.
We see that the probabilities $P_{1}(t)$
and $P_{2}(t)$ increase gradually from zero to a maximum value and
then decrease to zero as time increases. At first glance, the
maximum value of $P_{2}(t)$ is expected to decrease with the
increase of $g_{0}$ because the two-photon detuning $-2\nu$ (i.e., the
difference between the two-photon frequency shift $-4\nu$ and two
photon's total energy $-2\nu$) is proportional to $g_{0}^2$, and
therefore the photon blockade effect becomes stronger as $g_{0}$
increases. This trend is shown in the figure for $g_{0}/\omega_{M}$ from
$0.01$ to $0.6$. However, we see that the dependence of $P_{2}(t)$ on
$g_{0}$ is not monotonic due to the phonon sideband resonance effect.
In fact, Fig.~\ref{sinphoresonce}(c) shows that the photon blockade effect becomes
weaker when $g_{0}/\omega_{M}=1$. We can explain this feature by the
fact  when $g_{0}/\omega_{M}=1$, the two photons can induce the
resonant transitions
$|0\rangle_{a}|0\rangle_{b}\rightarrow|1\rangle_{a}|\tilde{0}(1)\rangle_{b}$
and $|1\rangle_{a}|\tilde{0}(1)\rangle_{b}
\rightarrow|2\rangle_{a}|\tilde{2}(2)\rangle_{b}$, and so the
maximum of $P_{2}(t)$ in this case is larger than that for
$g_{0}/\omega_{M}=0.6$.
%%%%%%%%%%%%%%%%%%%%%%%%%%%%%%%%%%%%%%%%%%%%%%%%%%%%
\begin{figure}[tbp]
\center
\includegraphics[bb=17 3 357 257, width=\columnwidth]{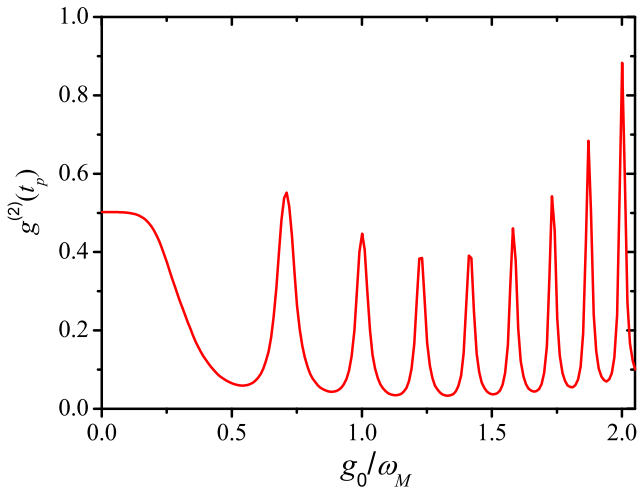}
\caption{(Color online) Plot of the equal-time second-order correlation
function  $g^{(2)}(t_{p})$ at time $\omega_{M}t_{p}=50$ vs the
scaled coupling strength $g_{0}/\omega_{M}$ in the single-photon
resonance case $\delta_{1}=\delta_{2}=-\nu$. Other parameters are
the same as those in Fig.~\ref{sinphoresonce}.}\label{g2vsganddelta}
\end{figure}
%%%%%%%%%%%%%%%%%%%%%%%%%%%%%%%%%%%%%%%%%%%%%%%%%%%%

The oscillating feature of the correlation function $g^{(2)}$ as a function of $g_{0}$ was
found in Ref.~\cite{Rabl2011}, which is based on an approximate
steady state solution of a continuously driven system. In our
case, our exact solution also indicates such a feature at
transient times. This is shown in Fig.~\ref{g2vsganddelta} in
which the dependence of $g^{(2)}(t)$ on $g_{0}$ at a given time
$t_{p}$ is plotted. Here $t_{p}$ is chosen when there are
appreciable amount of photons inside the cavity. We can see
clearly resonance peaks at specific values of $g_{0}$. To explain the
resonance, we note that when the frequencies of two photons are
$\delta_{1}=\delta_{2}=-\nu$ used in the figure, resonant
transitions $|0\rangle_{a}|0\rangle_{b}\rightarrow|2\rangle_{a}|\tilde{n}(2)\rangle_{b}$
are allowed if $-4\nu+n\omega_{M}=-2\nu$. This is the resonance
condition allowing two photons to exist in the cavity at the same
time. Specifically, the values of $g_{0}$ associated with these
resonance peaks are located at
\begin{equation}
\frac{g_{0}}{\omega_{M}}=\sqrt{\frac{n}{2}},\hspace{0.5
cm}n=0,1,2,\cdots,\label{phononsideresonance}
\end{equation}
which are consistent with the locations of the resonance peaks in
Fig.~\ref{g2vsganddelta}. We know from Eq.~(\ref{phononsideresonance}) that, with the increasing of $g_{0}$, the
sideband modulation peaks become more and more dense.

\begin{figure}[tbp]
\center
\includegraphics[bb=20 3 357 257, width=\columnwidth]{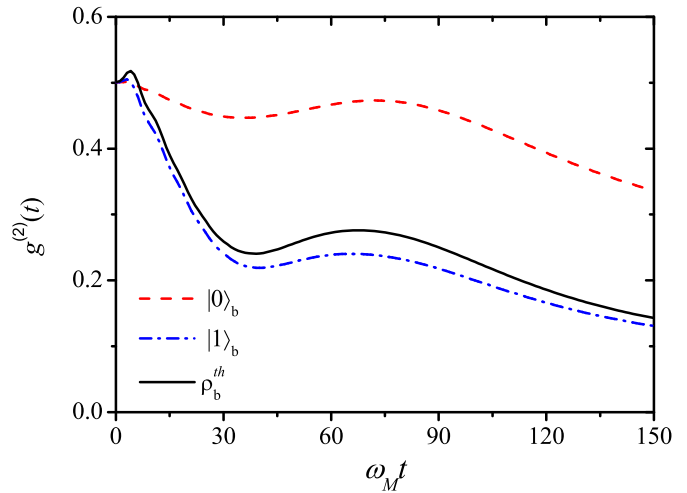}
\caption{(Color online) The second-order correlation function $g^{(2)}(t)$ vs the scaled time $\omega_{M}t$ when
the mirror's initial state is Fock state $|1\rangle_{b}$ (blue dot dashed curve) and thermal state $\rho^{th}_{b}(\bar{n}=1)$ (black solid curve).
The ground state $|0\rangle_{b}$ case (red dashed curve) is presented as a referrence. Other parameters are:
$g_{0}/\omega_{M}=0.3$, $\gamma_{c}/\omega_{M}=0.1$, $\epsilon/\omega_{M}=0.01$, and $\delta_{1}=\delta_{2}=-\nu$.}\label{mirrinistate}
\end{figure}

The above discussions have considered the initial ground state of the mirror.
It is interesting to ask how the $g^{(2)}(t)$ behaves when the mirror is in excited states initially.
Suppose the initial state of the mirror is $|n\rangle_{b}$, then the system can undergo the transitions
$|0\rangle_{a}|n\rangle_{b}\rightarrow|1\rangle_{a}|\tilde{n}(1)\rangle_{b}\rightarrow|2\rangle_{a}|\tilde{m}(2)\rangle_{b}$.
Therefore, the Franck-Condon factors $\langle n|_{b}|\tilde{n}(1)\rangle_{b}$ and $\langle \tilde{n}(1)|_{b}|\tilde{m}(2)\rangle_{b}$
are important to determine the magnitude of $g^{(2)}(t)$. In Fig.~\ref{mirrinistate}, we illustrate this
feature by considering a Fock state $|1\rangle_{b}$ as an initial state (blue dot dashed curve).
For the parameters used in this figure, the relatively small Franck-Condon factor
$\langle \tilde{1}(1)|_{b}|\tilde{1}(2)\rangle_{b}<\langle \tilde{0}(1)|_{b}|\tilde{0}(2)\rangle_{b}$ leads to a suppression of $P_{2}(t)$,
and hence  $g^{(2)}(t)$ can be substantially lower than that of the initial ground state (red dashed curve).
In Fig.~\ref{mirrinistate}, we have also plotted the $g^{(2)}(t)$ (black solid curve) for the initial thermal
state $\rho^{th}_{b}(\bar{n}=1)=\sum_{n_{0}=0}^{\infty}2^{-(n_{0}+1)}|n_{0}\rangle_{b}\langle n_{0}|_{b}$
($\bar{n}$ being the average thermal phonon number). We see that the $g^{(2)}(t)$ in this case is between those
in the cases of $|0\rangle_{b}$ and $|1\rangle_{b}$ due to statistical mixture.
%%%%%%%%%%%%%%%%%%%%%%%%%%%%%%%%%%%%%%%%%%%%%%%%%%%%
%%%%%%%%%%%%%%%%%%%%%%%%%%%%%%%%%%%%%%%%%%%%%%%%%%%%

\section{conclusion}

In conclusion, we have studied analytically the two-photon
scattering in a cavity optomechanical system. Under the
Wigner-Weisskopf framework, we have obtained the exact transient
solution of the system with the Laplace transform method. On one hand, the
long-time solution reveals the detailed physical transitions for the two
photons in the scattering process. This could help us to understand the
physical mechanism to induce two-photon correlation. In addition, from the poles of these scattering amplitudes, we
have obtained the resonance conditions of these transitions. In particular, the final state
of the two photons is generally correlated in frequency space
because of the frequency sum appearing in the denominator of the
amplitudes $C_{\textrm{III}}$ and $C_{\textrm{IV}}$. Based on the long-time solution,
we have calculated the two-photon joint spectrum, which shows clear evidence for two-photon frequency anti-correlation.
On the other hand, the transient dynamics of the equal-time second-order
correlation function of the cavity photons has been
calculated in order to address the photon blockade effect. We have found that the
photon blockade effect can be induced by the optomechanical coupling.
Besides, the correlation function has also been found to exhibit resonance peaks as a
function of the optomechanical coupling strength, and we have determined
the peak positions in Eq.~(\ref{phononsideresonance}) by a two-photon resonance condition.

\emph{Note added}. We notice a related paper that appeared recently~\cite{Liu2013}.

\begin{acknowledgments}
J. Q. Liao would like to thank J. F. Huang for technical
support. This work is partially supported by a grant from the
Research Grants Council of Hong Kong, Special Administrative Region
of China (Project No.~CUHK401810). J. Q. Liao is partially supported
by Japan Society for the Promotion of Science
(JSPS) Foreign Postdoctoral Fellowship No. P12503.
\end{acknowledgments}

\appendix*
\begin{widetext}

\section{Derivation of Eq.~(\ref{sctlontsolu})\label{apptwophoscat}}

In this Appendix, we give a detailed derivation of the two-photon
scattering solution in Eq.~(\ref{sctlontsolu}). By the Laplace
transform $\tilde{f}(s)=\int_{0}^{\infty}f(t) e^{-st}dt$,
Eq.~(\ref{equationofmotion}) becomes
\begin{subequations}
\begin{gather}
[s+i(m\omega_{M}-4\nu)]\tilde{A}_{m}(s)=A_{m}(0)-i\sqrt{2}\xi
\sum_{n=0}^{\infty}\int_{0}^{\infty}\langle\tilde{m}(2)\vert_{b}\vert
\tilde{n}(1)\rangle_{b}\tilde{B}_{n,k}(s)dk,\label{lapeq-am}\\
[s+i(\Delta_{k}-\nu +m\omega_{M})]\tilde{B}_{m,k}(s)=B_{m,k}(0)
-i\sqrt{2}\xi\sum_{n=0}^{\infty}\langle\tilde{m}(1)\vert_{b}\vert\tilde{n}(2)\rangle_{b}\tilde{A}_{n}(s)
-i\xi\sum_{n=0}^{\infty}\int_{0}^{\infty}\langle\tilde{m}(1)\vert_{b}\vert
n\rangle_{b}\tilde{C}_{n,p,k}(s)dp,\label{lapeq-bm}\\
[s+i(\Delta_{p}+\Delta_{q}+m\omega_{M})]\tilde{C}_{m,p,q}(s)=C_{m,p,q}(0)-i\xi\sum_{n=0}^{\infty}\langle
m\vert_{b}\vert\tilde{n}(1)\rangle_{b}[\tilde{B}_{n,p}(s)+\tilde{B}_{n,q}(s)],\label{lapeq-cm}
\end{gather}
\end{subequations}
where $A_{m}(0)$, $B_{m,k}(0)$, and $C_{m,p,q}(0)$ are the initial
conditions, which are given in Eq.~(\ref{scainicondition}). From
Eqs.~(\ref{lapeq-am}),~(\ref{lapeq-cm}), and
(\ref{scainicondition}), we have
\begin{subequations}
\begin{gather}
\tilde{A}_{m}(s)=\frac{-i\sqrt{2}\xi}{[s+i(m\omega_{M}-4\nu)]
}\sum_{n=0}^{\infty}\int_{0}^{\infty}\langle\tilde{m}(2)\vert_{b}\vert
\tilde{n}(1)\rangle_{b}\tilde{B}_{n,k^{\prime}}(s)dk^{\prime},\label{scatinfAm}\\
\tilde{C}_{m,p,q}(s)=\frac{1}{[s+i(\Delta_{p}+\Delta_{q}+m\omega_{M})]}\left(C_{m,p,q}(0)-i\xi\sum_{n=0}^{\infty}\langle
m\vert_{b}\vert\tilde{n}(1)\rangle_{b}[\tilde{B}_{n,p}(s)+\tilde{B}_{n,q}(s)]\right),\label{scatinfCm}
\end{gather}
\end{subequations}
Substitution of Eqs.~(\ref{scatinfAm}) and (\ref{scatinfCm}) into
Eq.~(\ref{lapeq-bm}) leads to
\begin{eqnarray}
&&\left[s+\frac{\gamma_{c}}{2}+i(\Delta_{k}-\nu +m\omega_{M})\right]\tilde{B}_{m,k}(s)\notag\\
&=&-\int_{0}^{\infty}\sum_{l,n=0}^{\infty}\left(\frac{2\xi
^{2}\langle \tilde{m}(1)\vert_{b}\vert\tilde{n}(2)\rangle_{b}\langle
\tilde{n}(2)\vert_{b}\vert\tilde{l}(1)\rangle_{b}}{[s+i(n\omega
_{M}-4\nu)]}+\frac{\xi^{2}\langle\tilde{m}(1)\vert_{b}\vert
n\rangle_{b}\langle n\vert_{b}\vert\tilde{l}(1)\rangle_{b}}{[s+i(
\Delta_{p}+\Delta_{k}+n\omega_{M})]}\right)\tilde{B}_{l,p}(s)dp\notag\\
&&+2i\pi\xi \mathcal{N}\langle\tilde{m}(1)\vert_{b}\vert
n_{0}\rangle_{b}\left(\frac{1}{[\Delta_{k}+\delta_{1}+n_{0}\omega_{M}-i(s+\epsilon)](\Delta_{k}-\delta_{2}+i\epsilon)}
+\delta_{1}\leftrightarrow\delta_{2}\right),
\end{eqnarray}
where $\gamma_{c}=2\pi\xi^{2}$ is the cavity-field decay rate.
We introduce a new variable $\tilde{F}_{m,k}(s)$ by
\begin{eqnarray}
\tilde{B}_{m,k}(s)&=&\frac{2\pi i\xi \mathcal{N}\langle\tilde{m}(1)\vert
_{b}\vert n_{0}\rangle_{b}}{[s+\frac{\gamma_{c}}{2}+i(\Delta
_{k}-\nu
+m\omega_{M})]}\left(\frac{1}{[\Delta_{k}+\delta_{1}+n_{0}\omega_{M}-i(
s+\epsilon)](\Delta_{k}-\delta_{2}+i\epsilon)}+\delta_{1}\leftrightarrow\delta_{2}\right)[1+\tilde{F}_{m,k}(s)].\label{scatrerFmintro}
\end{eqnarray}
Then the equation for $\tilde{F}_{m,k}(s)$ is obtained as
\begin{eqnarray}
&&\left(\frac{\langle\tilde{m}(1)\vert_{b}\vert
n_{0}\rangle_{b}}{[\Delta_{k}+\delta_{1}+n_{0}\omega_{M}-i(s+\epsilon)](\Delta_{k}-\delta_{2}+i\epsilon)}+\delta_{1}
\leftrightarrow\delta_{2}\right)\tilde{F}_{m,k}(s)\notag\\
&=&-\int_{0}^{\infty}\sum_{l,n=0}^{\infty}\left(\frac{2\xi^{2}\langle\tilde{m}(1)
\vert_{b}\vert\tilde{n}(2)\rangle_{b}\langle\tilde{n}(2)\vert
_{b}\vert\tilde{l}(1)\rangle_{b}}{[s+i(n\omega_{M}-4\nu)]}
+\frac{\xi^{2}\langle\tilde{m}(1)\vert_{b}\vert n\rangle_{b}\langle
n\vert_{b}\vert\tilde{l}(1)\rangle_{b}}{[s+i(\Delta
_{p}+\Delta_{k}+n\omega_{M})]}\right)\frac{\langle
\tilde{l}(1)\vert_{b}\vert
n_{0}\rangle_{b}}{[s+\frac{\gamma_{c}}{2}+i(\Delta_{p}-\nu+l\omega_{M})]}\nonumber\\
&&\times\left(\frac{1}{[\Delta_{p}+\delta_{1}+n_{0}\omega_{M}-i(s+\epsilon)](\Delta_{p}-\delta_{2}+i\epsilon
)}+\delta_{1}\leftrightarrow\delta_{2}\right)
\tilde{F}_{l,p}(s)dp+S_{1},\label{scaeqofFmk}
\end{eqnarray}
where
\begin{eqnarray}
S_{1}&=&\frac{1}{[\delta_{1}+\delta_{2}+n_{0}\omega_{M}-i(
s+2\epsilon)]}\sum_{l,n=0}^{\infty}\left[\left(\frac{2\gamma
_{c}\langle\tilde{m}(1)\vert_{b}\vert\tilde{n}(2)\rangle
_{b}\langle\tilde{n}(2)\vert_{b}\vert\tilde{l}(1)\rangle
_{b}\langle\tilde{l}(1)\vert_{b}\vert n_{0}\rangle_{b}}{[s+i(
n\omega_{M}-4\nu)][\delta_{1}-\nu+l\omega_{M}-i(s+\epsilon+\frac{\gamma
_{c}}{2})]}\right.\right.\nonumber\\
&&\left.\left.-\frac{i\gamma_{c}\langle\tilde{m}(1)\vert_{b}\vert
n\rangle_{b}\langle n\vert_{b}\vert
\tilde{l}(1)\rangle_{b}\langle\tilde{l}(1)\vert_{b}\vert
n_{0}\rangle_{b}}{[\delta_{1}+\Delta_{k}+n\omega_{M}-i(\epsilon+s)
][\delta_{1}-\nu+l\omega_{M}-i(s+\epsilon+\frac{\gamma_{c}}{2}
)]}\right)+(\delta_{1}\leftrightarrow\delta_{2})\right].
\end{eqnarray}
By introducing a new variable $x_{m}(s)$
\begin{eqnarray}
\tilde{F}_{m,k}(s)=\left(\frac{\langle\tilde{m}(1)\vert_{b}\vert
n_{0}\rangle_{b}}{(\Delta_{k}-\delta_{2}+i\epsilon)[\Delta_{k}+\delta_{1}+n_{0}\omega_{M}-i(s+\epsilon)]}
+\delta_{1}\leftrightarrow\delta_{2}\right)^{-1}
\left[S_{1}+x_{m}(s)\right],\label{scatrerxmintro}
\end{eqnarray}
we obtain the equation for $x_{m}(s)$ as follows:
\begin{eqnarray}
x_{m}(s)+\sum_{l,n=0}^{\infty}\frac{\gamma_{c}\langle\tilde{m}(1)\vert
_{b}\vert\tilde{n}(2)\rangle_{b}\langle\tilde{n}(2)\vert_{b}\vert
\tilde{l}(1)\rangle_{b}}{[s+i(n\omega_{M}-4\nu)]}x_{l}(s)
&=&-\sum_{l,n=0}^{\infty}\frac{2\gamma_{c}^{2}\langle\tilde{m}(1)\vert_{b}\vert
\tilde{n}(2)\rangle_{b}\langle\tilde{n}(2)\vert_{b}\vert\tilde{l}(1)\rangle_{b}\langle
\tilde{l}(1)\vert_{b}\vert
n_{0}\rangle_{b}}{[\delta_{1}+\delta_{2}+n_{0}\omega_{M}-i(s+2\epsilon
)][s+i(n\omega_{M}-4\nu)]^{2}}\nonumber\\
&&\times\left(\frac{1}{[\delta_{1}-\nu
+l\omega_{M}-i(s+\epsilon+\frac{\gamma_{c}}{2})]}+\delta_{1}\leftrightarrow\delta_{2}\right),\label{scaeqofxm}
\end{eqnarray}
The solution of Eq.~(\ref{scaeqofxm}) can be found as
\begin{eqnarray}
x_{m}(s)&=&\sum_{l,n=0}^{\infty}\frac{2\gamma_{c}^{2}}{[s+\gamma_{c}+i(
n\omega_{M}-4\nu)]}\frac{\langle\tilde{m}(1)\vert_{b}\vert
\tilde{n}(2)\rangle_{b}\langle\tilde{n}(2)\vert_{b}\vert\tilde{l
}(1)\rangle_{b}\langle\tilde{l}(1)\vert_{b}\vert
n_{0}\rangle_{b}}{[s+2\epsilon+i(\delta_{1}+\delta_{2}+n_{0}\omega_{M})]
[s+i(n\omega_{M}-4\nu)]}\nonumber\\
&&\times\left(\frac{1}{[(s+\epsilon+\frac{\gamma_{c}}{2})+i(\delta_{1}-\nu
+l\omega_{M})]}+\delta_{1}\leftrightarrow\delta_{2}\right).\label{scatxmsol}
\end{eqnarray}
Then by Eqs.~(\ref{scatinfAm}), (\ref{scatinfCm}),
(\ref{scatrerFmintro}), (\ref{scatrerxmintro}), and
(\ref{scatxmsol}), we can obtain
\begin{eqnarray}
\tilde{A}_{m}(s)& =&\frac{2\sqrt{2}\pi \mathcal{N}\gamma_{c}}{[s+2\epsilon+i(
\delta_{1}+\delta_{2}+n_{0}\omega_{M})][s+\gamma_{c}+i(m\omega_{M}-4\nu)]}\sum_{n=0}^{\infty}\left(\frac{\langle
\tilde{m}(2)\vert_{b}\vert
\tilde{n}(1)\rangle_{b}\langle\tilde{n}(1)\vert_{b}\vert
n_{0}\rangle_{b}}{[s+\epsilon+\frac{\gamma_{c}}{2}+i(\delta_{1}-\nu+n\omega_{M})]}
+\delta_{1}\leftrightarrow\delta_{2}\right),\label{srepofakscat}
\end{eqnarray}
\begin{eqnarray}
\tilde{B}_{m,k}(s)&=&-\frac{2\pi\xi \mathcal{N}\langle
\tilde{m}(1)|_{b}|n_{0}\rangle_{b}}{[s+\frac{\gamma
_{c}}{2}+i(\Delta_{k}-\nu +m\omega_{M})]}\left(\frac{1}{(\Delta
_{k}-\delta_{2}+i\epsilon)}\frac{1}{[s+\epsilon +i(\Delta
_{k}+\delta_{1}+n_{0}\omega_{M})]}+\delta_{1}\leftrightarrow\delta_{2}\right)\nonumber\\
&&-\sum_{l,n=0}^{\infty}\frac{4\pi i\xi \mathcal{N}\gamma_{c}\langle
\tilde{m}(1)|_{b}|\tilde{n}(2)\rangle_{b}\langle
\tilde{n}(2)|_{b}|\tilde{l}(1)\rangle_{b}\langle
\tilde{l}(1)|_{b}|n_{0}\rangle_{b}}{[s+\frac{\gamma
_{c}}{2}+i(\Delta_{k}-\nu+m\omega_{M})][s+2\epsilon+i(\delta_{1}+\delta_{2}+n_{0}\omega_{M})]}\nonumber\\
&&\times\frac{1}{[s+\gamma_{c}+i(n\omega_{M}-4\nu)]}\left(\frac{1}{[s+\epsilon+\frac{\gamma_{c}}{2}+i(\delta
_{1}-\nu +l\omega_{M})]}+\delta_{1}\leftrightarrow\delta_{2}\right)\nonumber\\
&&-\sum_{l,n}\frac{2\pi i\xi \mathcal{N}\gamma_{c}\langle
\tilde{m}(1)|_{b}|n\rangle_{b}\langle n|_{b}|\tilde{l}(1)\rangle
_{b}\langle\tilde{l}(1)|_{b}|n_{0}\rangle_{b}}{[s+\frac{\gamma
_{c}}{2}+i(\Delta_{k}-\nu+m\omega_{M})][s+2\epsilon+i(\delta_{1}+\delta_{2}+n_{0}\omega_{M})]}\nonumber\\
&&\times\left(\frac{1}{[s+\epsilon +i(\Delta_{k}+\delta
_{1}+n\omega_{M})][s+\epsilon+\frac{\gamma_{c}}{2}+i(\delta
_{1}-\nu+l\omega_{M})]}+\delta_{1}\leftrightarrow\delta_{2}\right),\label{srepofbmkscat}
\end{eqnarray}
\begin{eqnarray}
\tilde{C}_{m,p,q}(s)&=&\frac{\mathcal{N}}{[s+i(\Delta
_{p}+\Delta_{q}+m\omega_{M})]}\left[\frac{1}{(\Delta_{p}-\delta_{1}+i\epsilon
)}\frac{1}{(\Delta_{q}-\delta_{2}+i\epsilon)}\delta_{m,n_{0}}+\sum_{n=0}^{\infty}\frac{i\gamma
_{c}\langle m\vert_{b}\vert\tilde{n}(1)\rangle_{b}\langle
\tilde{n}(1)\vert_{b}\vert n_{0}\rangle_{b}}{[s+\frac{\gamma
_{c}}{2}+i(\Delta_{p}-\nu+n\omega_{M})]}\right.\nonumber\\
&&\left.\times\left(\frac{1}{[s+\epsilon
+i(\Delta_{p}+\delta_{1}+n_{0}\omega_{M})]}\frac{1}{(\Delta_{p}-\delta_{2}+i\epsilon)}+\delta_{1}\leftrightarrow\delta_{2}\right)\right.\nonumber\\
&&\left.-\sum_{n,n^{\prime},l=0}^{\infty}\frac{\gamma_{c}^{2}\langle
m\vert_{b}\vert\tilde{n}(1)\rangle_{b}\langle\tilde{n}(1)\vert_{b}\vert
n^{\prime}\rangle_{b}\langle n^{\prime}\vert_{b}\vert
\tilde{l}(1)\rangle_{b}\langle \tilde{l}(1)\vert_{b}\vert
n_{0}\rangle_{b}}{[s+2\epsilon +i(\delta_{1}+\delta_{2}+n_{0}\omega_{M})][s+\frac{\gamma_{c}}{2}+i(\Delta_{p}-\nu+n\omega_{M})]}\right.\nonumber\\
&&\left.\times\left(\frac{1}{[s+\epsilon+\frac{\gamma_{c}}{2}+i(\delta_{1}-\nu+l\omega_{M})]
}\frac{1}{[s+\epsilon+i(\Delta_{p}+\delta_{1}+n^{\prime}\omega_{M})]}+\delta_{1}\leftrightarrow\delta_{2}\right)\right.\nonumber\\
&&\left.-\sum_{n,n^{\prime},l=0}^{\infty}\frac{2\gamma_{c}^{2}\langle
m\vert_{b}\vert\tilde{n}(1)\rangle_{b}\langle
\tilde{n}(1)\vert_{b}\vert\tilde{n}^{\prime}(2)\rangle_{b}\langle
\tilde{n}^{\prime}(2)\vert_{b}\vert\tilde{l}(1)\rangle
_{b}\langle\tilde{l}(1)\vert_{b}\vert
n_{0}\rangle_{b}}{[s+2\epsilon+i(\delta_{1}+\delta_{2}+n_{0}\omega_{M})][s+\gamma
_{c}+i(n^{\prime}\omega_{M}-4\nu)][s+\frac{\gamma_{c}}{2}+i(\Delta_{p}-\nu+n\omega_{M})]}\right.\nonumber\\
&&\left.\times\left(\frac{1}{[s+\epsilon
+\frac{\gamma_{c}}{2}+i(\delta_{1}-\nu+l\omega_{M})]}+\delta_{1}\leftrightarrow\delta_{2}\right)\right]+\Delta_{p}\leftrightarrow
\Delta_{q}.
\end{eqnarray}
The transient solution of these probability amplitudes $A_{n_{0},m}(t)$,
$B_{n_{0},m,k}(t)$, and $C_{n_{0},m,p,q}(t)$ can be obtained by the inverse
Laplace transform. Here we add the subscript $n_{0}$ in the transient solution to mark the
mirror's initial state $|n_{0}\rangle_{b}$. The corresponding long-time solution is given in
Eq.~(\ref{sctlontsolu}).
\end{widetext}

\end{document}